\documentstyle[aps,prd,preprint,epsf]{revtex}
\newcommand{\tr}{\,{\rm tr}}
\newcommand{\vp}{\varphi}

\begin{document}
\setcounter{footnote}{1} \draft
\title{Mass Spectra of N=2 Supersymmetric SU(n) Chern-Simons-Higgs Theories}
\author{Hsien-chung Kao \footnote{Email address:hckao@mail.tku.edu.tw}}
\address{Academia Sinica, Institute of Physics, Taipei, Taiwan, R.O.C. \\ {\rm and} \\
Department of Physics, Tamkang University, Tamsui, Taiwan, R.O.C. \footnote{Permanent
address.}}


\maketitle
\begin{abstract}
An algebraic method is used to work out the mass spectra and symmetry breaking patterns
of general vacuum states in $N=2$ supersymmetric $SU(n)$ Chern-Simons-Higgs systems with
the matter fields being in the adjoint representation.  The approach provides with us a
natural basis for fields, which will be useful for further studies in the self-dual
solutions and quantum corrections.  As the vacuum states satisfy the $SU(2)$ algebra, it
is not surprising to find that their spectra are closely related to that of angular
momentum addition in quantum mechanics.  The analysis can be easily generalized to other
classical Lie groups.
\end{abstract}

Chern-Simons-Higgs systems possess many unusual features and have been extensively
studied in the past. For a review, see Ref. \cite{CSReview}. First, the Abelian theories
contain particle excitations carrying fractional spin and statistics, which can be used
to describe the fractional quantum Hall effect \cite{CSFT,CSANYON,FQHE}. Furthermore,
relativistic Chern-Simons-Higgs theories become self-dual when the Higgs potential takes
on a special sixth order form. In such case, the energy functional of the systems has a
Bogomol'nyi bound, which is saturated by the solutions of the self-dual equations
\cite{Bogo,Hong}. There are both topological solitons and nontopological vortices
\cite{Jackiw}. Similar self-dual structure has also been found in non-Abelian
Chern-Simons-Higgs systems with an additional global $U(1)$ symmetry \cite{klee1}. When
the Higgs fields are in the fundamental representation, the non-Abelian self-dual
solitons are equivalent to the Abelian ones \cite{klee2}. When the Higgs fields are in
the adjoint representation, the analysis of the self-dual equations is generally much
more involved.  There are many inequivalent degenerate vacua exhibiting different
patterns of symmetry breaking. As the vacuum states satisfy an algebraic equation that is
equivalent to the $SU(2)$, they are classified by how the $SU(2)$ algebra is embedded in
a general Lie algebra \cite{Spectra}. In the $SU(2)$ case, the self-dual equations are
identical to the Abelian ones \cite{klee1}. The $SU(3)$ model is the simplest one with
full-fledged non-Abelian characteristics, and has also been studied in detail
\cite{NASD}. It is known that there is an underlying $N=2$ supersymmetry behind the
self-duality in these systems and thus the Bogomol'nyi bound is expected to be preserved
in the quantum regime \cite{CSII,Witten}. Moreover, when these self-dual theories are
dimensionally reduced, an additional Noether charge appears. As a result, a BPS-type of
domain wall is induced \cite{Domn}.

Another interesting aspect of the Chern-Simons-Higgs theories is the quantum correction
to the Chern-Simons coefficient.  In the Abelian case, it has been shown that only the
fermion one-loop diagram contributes to the correction to the Chern-Simons coefficient
and yields ${1\over 4\pi}$, if there are neither massless charged particles nor
spontaneous breaking of gauge symmetry \cite{Coleman}.   When there is spontaneous
symmetry breaking, by taking into account the effect of the would be Chern-Simons term,
it has been shown that the one-loop correction in the Higgs phase is identical to that in
the symmetric phase \cite{Khleb2,CHTh}. In the non-Abelian case, the Chern-Simons
coefficient must be integer multiple of ${1\over 4\pi}$ for the theories to be
quantum-mechanically consistent. Therefore, it is interesting to see whether the
quantization condition is spoiled by quantum effects.  In the symmetric phase, the
condition is preserved in one and two loop order \cite{Pisarski,YCKao}.  In the Higgs
phase, the situation is complicated by choices of gauge group and representation of the
Higgs fields. When the Higgs fields are in fundamental $SU(n)$ or $SO(n)$, the
quantization condition is shown to hold for all $n$ by carefully subtracting out the
contribution of the would be Chern-Simons term \cite{Khare,Nahill}. When the Higgs fields
are in the adjoint representation, only partial calculations have been done and the
results are perplexing \cite{Dunne}.  Supersymmetry also imposes interesting properties
on quantum corrections of the Chern-Simons coefficient.  For $N=2,3$ Yang-Mills
Chern-Simons theories, the one-loop quantum corrections vanishes because of the
cancellation between the bosonic and fermionic contributions  \cite{SUSYYMCS}. As for
$N=2,3$ Chern-Simons-Higgs theories, the corrections are identical in the symmetric and
Higgs phases, reflecting that supersymmetry is preserved  \cite{SUSYCSH}.

In this paper, we use an algebraic method to work out the mass spectra and symmetry
breaking patterns of general vacuum states in $N=2$ supersymmetric $SU(n)$
Chern-Simons-Higgs systems with the matter fields being in the adjoint representation.
Supersymmetry is helpful in clarifying some misinterpretation of the mass spectra.
Although part of the results have been reported \cite{Spectra}, our approach is more
powerful and provides with us a natural basis for fields, which will also be useful for
further studies in the self-dual solutions and quantum corrections.  As the vacuum states
satisfy the $SU(2)$ algebra, it is not surprising to find that their spectra are closely
related to the pattern of angular momentum addition in quantum mechanics.

With the Higgs fields in the adjoint representation, the $N=2$ supersymmetric non-Abelian
Chern-Simons Higgs theories are given by\cite{SUSYIII}:
\begin{eqnarray}
{\cal L} = \tr  \biggl\{ &\;& \kappa \; \epsilon^{\mu\nu\rho}\left( A_\mu \partial_\nu
A_\rho - {2\over 3}iA_\mu A_\nu A_\rho \right)
 + 2 |D_\mu \Phi|^2  + 2i  \bar{\psi} D \hskip -0.65em \slash\, \psi
 -  {2v^2\over \kappa} \bar{\psi} \psi \nonumber \\
&\;& - {2\over \kappa^2} \left | [\Phi, [\Phi^\dagger, \Phi]] - v^2 \Phi \right |^2 +
{2\over \kappa}[\Phi^\dagger, \Phi] [\bar{\psi},\psi] + {4\over \kappa} [\bar{\psi},
\Phi] [\Phi^\dagger,  \psi] \biggr\}, \label{L_full} \\  \nonumber
\end{eqnarray}
where  $A_\mu = A^a_\mu T^a$, $\Phi = \Phi^a T^a$, $\psi = \psi^a T^a$ and $D_\mu =
\partial_\mu - i [A_\mu, ~]$. We choose $\gamma_0= \sigma_2$, $\gamma_1= -i\sigma_3$,
$\gamma_2= -i\sigma_1$ so that $\gamma_\mu \gamma_\nu = g_{\mu\nu} -
i\epsilon_{\mu\nu\rho} \gamma^\rho$. Here, $g_{\mu\nu}={\rm diag}\left(1,-1,-1 \right)$
and $\epsilon^{012} =1$. The generators of the gauge group satisfy $[T^a, T^b] =
if^{abc}T^c,$ with the normalization $\tr\{T^a T^b\} = \frac{1}{2}\delta^{ab}$.

Physical states obey the constraint of Gauss's law:
\begin{eqnarray}
{\cal G}  \equiv  \kappa F_{12} + i[D_0 \Phi^\dagger, \Phi] - i[\Phi^\dagger, D_0\Phi] -
\{\psi^\dagger, \psi\}= 0,
\end{eqnarray}
which is obtained from the variation of $A_0$ and generates the local gauge
transformation. There exists in the system an additional global $U(1)$ symmetry whose
charge is $Q = \int d^2r\; 2\tr \left\{i\Phi^\dagger(D_0\Phi) - i(D_0\Phi)^\dagger \Phi +
\psi^\dagger\psi  \right\}$. Furthermore, it is also invariant up to a total derivative
under the following $N=2$ supersymmetric transformation:
\begin{eqnarray}
\delta A_\mu & = & {1\over\kappa}\bar{\alpha} \gamma_\mu [\Phi^\dagger, \psi] +
{1\over\kappa}[\bar{\psi}, \Phi] \gamma_\mu \alpha,  \nonumber \\ \delta \Phi & = &
-\bar{\alpha} \psi, \\ \delta \psi & = & \left\{i D \hskip -0.65em \slash\, \Phi -
{1\over\kappa} X \right\}\alpha, \nonumber
\end{eqnarray}
where $X \equiv \left([ \Phi, [\Phi^\dagger, \Phi]]- v^2 \Phi \right)$. The corresponding
supercharge is given by
\begin{eqnarray}
{\cal R} & = & \int d^2r\; 2\tr \left\{ \left(D \hskip -0.65em \slash\,\Phi^\dagger -
{i\over\kappa}X^\dagger  \right)\gamma^0 \psi \right\}.
\end{eqnarray}
It satisfies an $N=2$ superalgebra with a central charge:
\begin{equation}
\left[\bar{\alpha}{\cal R}, \bar{\cal R} \beta \right] = \bar{\alpha} \gamma^\mu \beta
P_\mu - \bar{\alpha} \beta Z, \label{super}
\end{equation}
with $Z = \frac{v^2}{\kappa} Q$. Multiply ${1\over 2}(1 \pm \gamma^0)$ to the
superalgebra in Eq. (\ref{super}) and take trace.  We then have
\begin{eqnarray}
{\cal P}^0 = \mp Z + \sum_A \{({\cal R}_\pm)_A, ({\cal R}_\pm^\dagger)_A \}
\end{eqnarray}
after making use of Gauss's law.  Here,
\begin{eqnarray}
{\cal R}_\pm &\equiv& {1 \pm \gamma^0 \over 2} {\cal R} \nonumber
\\ &=& \int d^2r \tr \biggl\{ (1 \pm \gamma^0)
\psi(D_0\Phi^\dagger \mp {i\over\kappa}X^\dagger)  \pm (\gamma^1 \pm i\gamma^2)\psi(D_1
\mp iD_2) \Phi^\dagger \biggr\}.
\end{eqnarray}
The expectation value of the energy has a Bogomol'nyi bound:
\begin{eqnarray}
\langle {\cal P}^0 \rangle \ge |\langle Z \rangle|. \nonumber
\end{eqnarray}
It is saturated when the following self-dual equations are satisfied.
\begin{eqnarray}
&\;& D_0\Phi \pm {i\over\kappa}([\Phi, [\Phi^\dagger, \Phi]]- v^2 \Phi) = 0 \nonumber
\\ &\;& (D_1 \pm iD_2) \Phi =0.
\end{eqnarray}

The vacuum satisfy
\begin{equation}
[\Phi_0, [\Phi^\dagger_0, \Phi_0]] - v^2 \Phi_0 = 0.
\end{equation}
Define $\Phi_0 = v\vp$, and we have $[\vp, [\vp^\dagger, \vp]] - \vp = 0.$   This is
nothing but the algebra of $SU(2)$ with $\vp = \frac{1}{\sqrt{2}} J_+$, and $[\vp,
\vp^\dagger] = J_z$.  As a result, the vacuum are classified by how the $SU(2)$ is
embedded into a general Lie algebra \cite{Spectra}.  It is well-known for $SU(2)$ that in
each given dimension there is only one inequivalent irreducible representation from which
one can construct arbitrary finite dimensional representations. Therefore, when the gauge
group is $SU(n)$, the number of inequivalent vacua is equal to $p(n)$, the number of
integer partitions of $n$. More explicitly, it can be arranged that $n = \sum_{i=1}^{k}
m_i n_i$ with $n_1 > n_2> \ldots > n_k$ without loss of generality.  Here, $n_i$ is the
dimensionality of the irreducible representation and $m_i$'s (the multiplicities) are
positive integers. For example, in the symmetric phase $k=1$ and $(m_1,n_1) = (n,1)$. In
such case, $(\Phi, \psi)$ saturate the Bogomol'nyi bound and form a reduced
representation of the $N=2$ superalgebra. The other extreme is the maximal embedding
case, $k=1$ and $(m_1,n_1) = (1,n)$.  Since it is also an irreducible representation of
$SU(2)$ and forms a building block for representations in higher dimensions, let us
concentrate on this case first. To quadratic term, the gauge part of the Lagrangian is
given by
\begin{eqnarray}
({\cal L}_0)_{gauge} = \tr  \biggl\{ \kappa \; \epsilon^{\mu\nu\rho} A_\mu \partial_\nu
A_\rho
 - 2v^2 A^\mu [\vp^\dagger, [\vp, A_\mu]] \biggr\}.
\label{L0_gauge}
\end{eqnarray}

To work out the spectrum, let us consider a "state" $\hat{\alpha}$ satisfying the
following equation
\begin{equation}
[\vp^\dagger, [\vp, \hat{\alpha}]] = \lambda\hat{\alpha}.
\end{equation}
Making use of the Jacobi identity, one can show that the operator, $[\vp, [\vp^\dagger,
~]]$, commutes with $[[\vp, \vp^\dagger],~]$. Therefore, we can choose $\hat{\alpha}$ to
be an simultaneous "eigenstate" of them, i.e.
\begin{equation}
[[\vp, \vp^\dagger] , \hat{\alpha}] = m \hat{\alpha}.
\end{equation}
The two "eigenvalues" $(\lambda, m)$ can be used to characterize $\hat{\alpha}$.  $m$ is
the "eigenvalue" of $J_z$, which will always be an integer as we will see later.  Further
use of the Jacobi identity reveals that $[\vp, \hat{\alpha}]$ and $[\vp^\dagger,
\hat{\alpha}]$ are again "eigenvectors" of the two operators, with eigenvalues
$(\lambda-m-1, m+1)$ and $(\lambda+m, m-1)$.  Moreover since $[\vp, \vp^j] = 0$,
$\vp^j$'s are like the highest weight states with "eigenvalues" $(0, j)$, where
$j=1,\ldots, n-1$. Thus, we can construct from $\vp^j$ a series of descending
"eigenstates", $[\vp^\dagger, \vp^j]_{(k)}$, with $k=0, \dots, 2j$.  Here, $[\vp^\dagger,
\vp^j]_{(k)}= [\vp^\dagger, [\vp^\dagger, \vp^j]_{(k-1)}]$, and $[\vp^\dagger,
\vp^j]_{(0)}= \vp^j$.  It is obvious that $k = j - m$. From now on, we will use $(j,m)$
to classify the "eigenstates" for convenience. Since their ranges are $j= 1,\ldots, n-1$,
and $m = -j, \ldots, j$, there are $n^2-1$ "eigenstates" in total. It can be shown that
states with different "eigenvalues" $(j,m)$ are orthogonal to one another. Hence, we can
also use them as a basis of the $SU(n)$ algebra. For consistency, we will choose the
normalization $\tr\left\{\hat{\alpha}^\dagger(j,m) \hat{\alpha}(j',m')\right\} =
\frac{1}{2} \delta_{j,j'} \delta_{m,m'}$ like the generators $T^a$'s.   In terms of
$(j,m)$, we now have
\begin{equation}
\lambda(j,m) = \lambda(j,m+1) + m+1,
\end{equation}
with $\lambda(j,j) = 0$, and
\begin{eqnarray}
&\;& [\vp^\dagger, [\vp , \hat{\alpha}(j,m)]] = \lambda(j,m) \hat{\alpha}(j,m),  \\ &\;&
[[\vp, \vp^\dagger], \hat{\alpha}(j,m)] = m \hat{\alpha}(j,m), \label{charge}
\end{eqnarray}
Solving the iteration equation for $\lambda(j,m)$, we see $\lambda(j,m) = \frac{1
}{2}\{j(j+1)-m(m+1)\}$. This is hardly surprising from hind sight, since $\frac{1}{2} J_-
J_+= \frac{1}{2}(J^2 - J^2_z - J_z)$. Furthermore, the effects of $\vp$ and $\vp^\dagger$
on the state $\hat{\alpha}(j,m)$ can be summarized as
\begin{eqnarray}
&\;& [\vp,  \hat{\alpha}(j,m)] = C_+(j,m) \hat{\alpha}(j,m+1), \\ &\;& [\vp^\dagger,
\hat{\alpha}(j,m)] = C_-(j,m) \hat{\alpha}(j,m-1). \nonumber
\end{eqnarray}
From the first equation, we see immediately
\begin{eqnarray}
\tr \left\{[\hat{\alpha}^\dagger(j,m), \vp^\dagger][\vp, \hat{\alpha}(j,m)] \right\} =
\frac{1}{2}|C_+(j,m)|^2 = \frac{1}{2}\lambda(j,m). \nonumber
\end{eqnarray}
Making use of the Jacobi identity, one can further determine the coefficients $C_-(j,m)$:
\begin{eqnarray}
C_-(j,m) = \sqrt{\lambda(j,m)+m}= \sqrt{\lambda(j,m-1)}.
\end{eqnarray}
Thus, it is just like the $SU(2)$ algebra.

To find the mass spectrum of the gauge field, all we have to do now is expand $A_\mu$ in
terms of $\hat{\alpha}(j,m)$'s: $A_\mu = A_\mu(j,m) \hat{\alpha}(j,m)$.  $A_\mu$ is
hermitian and $\hat{\alpha}(j,-m) = (-1)^m \hat{\alpha}^\dagger(j,m)$.  Thus,
$A_\mu(j,-m) = (-1)^m A^*_\mu(j,m)$, and $A_\mu(j,0)$ is real.  By making use of the
expansion, Eq. (\ref{L0_gauge}) becomes
\begin{eqnarray}
({\cal L}_0)_{gauge} = &\;& \quad \sum_{j=0}^{n-1} \biggl\{ \frac{1}{2}\kappa
\epsilon^{\mu\nu\rho} A_\mu(j,0) \partial_\nu A_\rho(j,0) - \frac{1}{2}j(j+1)\kappa M
A^2_\mu(j,0) \biggr\} \\ &\;& + \sum_{j=0}^{n-1} \sum_{m=1}^{j}  \left\{\kappa
\epsilon^{\mu\nu\rho} A^*_\mu(j,m) \partial_\nu A_\rho(j,m) - \{j(j+1)-m^2\} \kappa M
|A_\mu(j,m)|^2 \right\},\nonumber
\end{eqnarray}
where $M = v^2/\kappa$.  From the above results, we see every component of the gauge
field acquires a mass term.  Therefore, the $SU(n)$ gauge symmetry is completely broken
down.  Since $[[\vp, \vp^\dagger], \vp] = \vp$, there is a remaining global $U(1)$
symmetry whose generator is given by $\tilde{Q} = Q + 2\tr \{ {\cal G} [\vp,
\vp^\dagger]\}$.  It is obvious that $A_\mu(j,m)$ carries $m$ unit of $\tilde{Q}$-charge
because of Eq. (\ref{charge}).

To find the mass spectra of the Higgs field, we let $\Phi = \phi + v\vp$ and collect
terms quadratic in $\phi$.
\begin{eqnarray}
({\cal L}_0)_{Higgs} = \tr  \biggl\{ 2 |\partial_\mu \phi|^2 - 2 M^2\left | [\phi,
[\vp^\dagger, \vp]] + [\vp, [\vp^\dagger, \phi]]+ [\vp, [\phi^\dagger, \vp]] - \phi
\right |^2 \biggr\}. \label{L0_Higgs}
\end{eqnarray}
Expanding $\phi$ in terms of the basis $\hat{\alpha}(j,m)$, we see
\begin{eqnarray}
&\;& [\phi, [\vp^\dagger, \vp]] + [\vp, [\vp^\dagger, \phi]]+ [\vp, [\phi^\dagger, \vp]]
- \phi \nonumber\\ &\;& = \sum_{j=1}^{n-1} \sum_{m=-j}^{j} \lambda(j,m-2) \phi(j,m)
\hat{\alpha}(j,m) - \sum_{j=1}^{n-1} \sum_{m=-j+2}^{j} (-1)^m
\sqrt{\lambda(j,m-2)\lambda(j,m-1)} \phi^*(j,2-m) \hat{\alpha}(j,m) \nonumber \\ &\;& =
\quad \sum_{j=1}^{n-1} \{-(j+1)\} \phi(j,-j) \hat{\alpha}(j,-j) \\ &\;& \quad -
\sum_{j=1}^{n-1} \sum_{m=-j+2}^{j} \left\{ \lambda(j,m-2) -(-1)^m
\sqrt{\lambda(j,m-2)\lambda(j,m-1)} \phi^*(j,2-m) \right\} \hat{\alpha}(j,m). \nonumber
\end{eqnarray}
Consequently,
\begin{eqnarray}
&\;& ({\cal L}_0)_{Higgs} \nonumber\\ &\;& = \quad \sum_{j=1}^{n-1} \left\{|\partial_\mu
\phi(j,-j)|^2-(j+1)^2 |\phi(j,-j)|^2 + \frac{1}{2}[\partial_\mu
\tilde{\phi}(j,0)]^2-\frac{1}{2}\{j(j+1)\}^2 \tilde{\phi}^2(j,0) \right\} \\ &\;& \quad +
\sum_{j=1}^{n-1} \sum_{m=1}^{j-1} \left\{|\partial_\mu
\tilde{\phi}(j,m)|^2-\{j(j+1)-m^2\}^2 |\tilde{\phi}(j,m)|^2 \right\}. \nonumber
\end{eqnarray}
Here,
\begin{eqnarray}
\tilde{\phi}(j,m) = \sqrt{\frac{\lambda(j,m-1)}{j(j+1)-m^2}}\phi(j,m+1) +(-1)^m
\sqrt{\frac{\lambda(j,m)}{j(j+1)-m^2}} \phi^*(j,-m+1), \nonumber
\end{eqnarray}
for $m=0,\ldots,j-1$. In particular, $\tilde{\phi}(j,0) = \sqrt{\frac{1}{2}}\{\phi(j,1) +
\phi^*(j,1)\}$ is real. Note that $\tilde{\phi}(j,m)$ and $\phi^*(j,-j)$ carry $m$ and
$(j+1)$ units of $\tilde{Q}$-charges, respectively.

Now, let us pause to discuss the topology of a maximal embedding vacuum.  It is known
that the ground state can always be brought into the standard form by a gauge
transformation
\begin{eqnarray}
\Phi_0 = \left(\matrix{ 0 &c_1 &0 &0 &\ldots &0 \cr 0 &0 &c_2 &0 &\ldots &0 \cr 0 &0 &0
&c_3 &0  &\vdots \cr \vdots &\vdots &\ddots &\ddots &\ddots &\vdots \cr 0 &\vdots &\ldots
&\ddots &0 &c_{n-1} \cr 0 &0 &\ldots &\ldots &0 &0 }\right).
\end{eqnarray}
Here, $c_1, c_2, \ldots,$ and $c_{n-1}$ are all non-zero. The remaining gauge
transformations that leave $\Phi_0$ in the above standard form can be described uniquely
by
\begin{eqnarray}
U = \left(\matrix{ {\rm e}^{i 2\pi \theta_1} &0 &0 &\ldots &0 \cr 0 &{\rm e}^{i 2\pi
\theta_2} &0 &\ldots &0 \cr 0 &0 &{\rm e}^{i 2\pi \theta_3}  &0  &\vdots \cr \vdots
&\vdots &\ddots &\ddots &0 \cr 0 &0 &\ldots &0 &{\rm e}^{-i2\pi(\theta_1 +\ldots
+\theta_{n-1})} }\right),
\end{eqnarray}
with $0 \le \theta_i < 1$ for $i=1,\ldots, n-1$. Under such gauge transformation
\begin{eqnarray}
\Phi_0 \to U \Phi U^\dagger = \left(\matrix{ 0 &c_1 {\rm e}^{i 2\pi (\theta_1 -
\theta_2)} &0 &0 &\ldots &0 \cr 0 &0 &c_2 {\rm e}^{i 2\pi (\theta_2 - \theta_3)}&0
&\ldots &0 \cr 0 &0 &0 &c_3 {\rm e}^{i 2\pi (\theta_3 - \theta_4)}&0 &\vdots \cr \vdots
&\vdots &\ddots &\ddots &\ddots &\vdots \cr 0 &\vdots &\ldots &\ddots &0 &c_{n-1} {\rm
e}^{i 2\pi (\theta_1 +\ldots +\theta_{n-2} +2\theta_{n-1})} \cr 0 &0 &\ldots &\ldots &0
&0 }\right).
\end{eqnarray}
Therefore, $U$ induces in the parameter space $(\theta_1, \ldots, \theta_{n-1})$ a linear
transformation
\begin{eqnarray}
\left(\matrix{ \theta_1 \cr \theta_2 \cr \vdots \cr \theta_{n-2} \cr \theta_{n-1}
}\right) \to \left(\matrix{ 1 &-1 &0 &\ldots &0 \cr 0 &1 &-1 &\ddots &\vdots \cr \vdots
&\ddots &\ddots &\ddots &0 \cr  0 &\ldots &0 &1 &-1 & \cr 1 &1 &\ldots &1 &2 }\right)
\left(\matrix{ \theta_1 \cr \theta_2 \cr \vdots \cr \theta_{n-2} \cr \theta_{n-1}
}\right) .
\end{eqnarray}
It is easy to show that the determinant of the above linear transformation is $n$, and
the topology of $\Phi_0$ is thus $Z_n$.

Similarly, we can find the mass spectra of the fermionic part by expanding $\psi$ in the
$\hat{\alpha}(j,m)$ basis:
\begin{eqnarray}
({\cal L}_0)_{Fermion} &\;& = \tr  \biggl\{ 2i \bar{\psi}\partial \hskip -0.60em \slash\,
\psi - 2M \bar{\psi} \psi + 2M [\vp^\dagger, \vp] [\bar{\psi},\psi] + 4M [\bar{\psi},
\vp] [\vp^\dagger,  \psi] \biggr\}, \nonumber \\ &\;& = \sum_{j=1}^{n-1} \sum_{m= -j}^{j}
\left[ i\bar{\psi}(j,m)\partial \hskip -0.60em \slash\, \psi(j,m) + \{ j(j+1)- (m-1)^2 \}
M \bar{\psi}(j,m) \psi(j,m) \right] \nonumber \\ &\;& = \sum_{j=1}^{n-1} \biggl[
i\bar{\psi}(j,-j)\partial \hskip -0.60em \slash\, \psi(j,-j) - \{ j+1\} M
\bar{\psi}(j,-j) \psi(j,-j) \nonumber \\ &\;& \qquad +\, i\bar{\psi}(j,-j+1)\partial
\hskip -0.60em \slash\, \psi(j,-j+1) + \{ j \} M \bar{\psi}(j,-j+1) \psi(j,-j+1) \biggr]
\\ &\;& + \sum_{j=1}^{n-1}\sum_{m= -j+1}^{j-1} \biggl[ i\bar{\psi}(j,m+1)\partial \hskip
-0.60em \slash\, \psi(j,m+1) +\, \{ j(j+1)- m^2 \} M \bar{\psi}(j,m+1) \psi(j,m+1)
\biggr]. \nonumber \label{L0_Fermion}
\end{eqnarray}
Likewise, $\psi^*(j,m+1)$ carries $m$ units of $\tilde{Q}$-charge.

For convenience, we will refer the above spectrum as the $SU(n)$ one. Unlike the case of
fundamental matter coupling, some of the Higgs fields, $\phi(j,-j)$, saturate the
Bogomol'nyi bound. Indeed, it can be seen clearly from the tables that
$\left\{A_\mu(j,j), \psi^*(j,-j+1)\right\}$ and $\left\{\psi(j,-j),\phi(j,-j)\right\}$
form reduced representations of the $N=2$ supersymmetry separately. In the literature,
$A_\mu(j+1,j+1)$ and $\phi(j,-j)$ for $j=1,2, \ldots, j-1$ are paired up because of
ostensible degeneracy \cite{Spectra}. However, since their super-partners $\psi(j,-j+1)$
and $\psi(j,-j)$ have opposite helicities, we known this can not be right.

To obtain the mass spectra of a general vacuum, let us proceed by considering the case
that it is formed by the direct sum of two maximal embeddings with dimensions $n_1$ and
$n_2$
\begin{eqnarray}
\vp = \left(\matrix{ \vp_1&0\cr 0&\vp_2}\right). \nonumber
\end{eqnarray}
It is obvious that in the "diagonal" sector we have two kinds of "eigenstates" $$
\left(\matrix{\hat{\alpha}_1(j_1,m_1)&0\cr 0&0}\right)\quad {\rm and}\quad
\left(\matrix{0&0\cr 0&\hat{\alpha}_2(j_2,m_2)}\right) $$ with "eigenvalues" $(j_1, m_1)$
and $(j_2,m_2)$. The number of generators in the two subspaces are $n_1^2-1$ and
$n_2^2-1$, respectively.

In the "off-diagonal" sector,
\begin{eqnarray}
[\vp, \left(\matrix{ 0&\Omega\cr 0&0}\right) ]
 =  \left(\matrix{
0&\vp_1 \Omega - \Omega \vp_2\cr 0&0}\right). \nonumber
\end{eqnarray}
It is well known that if $J_i$ satisfies the angular momentum algebra, so does $
\bar{J}_i \equiv  - J^*_i$.  In the usual representation we adopt for angular momentum,
$\bar{J}_\pm = - J_\pm$ and $\bar{J}_z = - J_z$.  Let us define $|j,m\rangle\!\rangle$ as
the simultaneous eigenstate of $\bar{J}^2$ and $\bar{J}_z$.  By choosing
$|j,j\rangle\!\rangle = |j,-j\rangle$, it is obvious $|j,m\rangle\!\rangle =
(-1)^{j-m}|j,-m\rangle$.  In terms of $|j_1,m_1\rangle$ and $|j_2,m_2\rangle\!\rangle$ we
can construct the total angular momentum eigenstates $$ \hat{\beta}(j,m) = \sum_{m1,m2}
|j_1,m_1\rangle \langle\!\langle j_2,m_2| \; \frac{1}{\sqrt{2}}\langle j1, m1;j2, m2|
j,m;j_1,j_2\rangle, $$ where $\langle j1, m1;j2, m2| j,m;j_1,j_2\rangle$ are the
well-known Clebash-Gordan coefficients.  Like $\hat{\alpha}(j,m)$, these states satisfy
\begin{eqnarray}
&\;& [\vp,  \hat{\beta}(j,m)] = \sqrt{\lambda(j,m)} \hat{\beta}(j,m+1), \\ \nonumber &\;&
[\vp^\dagger, \hat{\beta}(j,m)] = \sqrt{\lambda(j,m-1)}\hat{\beta}(j,m-1), \\ &\;&
[[\vp,\vp^\dagger], \hat{\beta}(j,m)] = m \hat{\beta}(j,m). \nonumber
\end{eqnarray}
Together with $\hat{\beta}^\dagger(j,m)$, we see there are $2n_1 n_2$ generators in the
"off-diagonal" sector. \vskip0.6cm

The last state is given by $\hat{\alpha}_0 = \left(\matrix{ \sqrt{\frac{n_2}{2n_1
n}}I_1&0\cr 0&-\sqrt{\frac{n_1}{2n_2 n}}I_2}\right)$, with $I_1$ and $I_2$ the identity
\linebreak \vskip0.2cm \noindent operators in the two subspace. It commute with both
$\hat{\alpha}_1(j_1,m_1)$ and $\hat{\alpha}_2(j_2,m_2)$.

Now we are ready to expand $A_\mu$:
\begin{eqnarray}
A_\mu &\;& = (A_1)_\mu(j_1,m_1) \hat{\alpha}_1(j_1,m_1) + (A_2)_\mu(j_2,m_2)
\hat{\alpha}_2(j_2,m_2) +(A_0)_\mu \hat{\alpha}_0 \\ \nonumber &\;& \quad +
\sum_{j=\frac{1}{2}|n_1-n_2|}^{\frac{1}{2}(n_1+n_2)-1}\; \sum_{m=-j}^{j}B_\mu(j,m)
\hat{\beta}(j,m) + B^*_\mu(j,m) \hat{\beta}^\dagger(j,m) . \nonumber
\end{eqnarray}
Note that $(A_0)_\mu$ is real since $A_\mu$ is hermitian.  The mass spectra for
$(A_1)_\mu$ and $(A_2)_\mu$ are already known, and we will just give the part for
$(A_0)_\mu$ and $B_\mu$
\begin{eqnarray}
({\cal L}'_0)_{gauge} = &\;& \frac{ \kappa}{2} \epsilon^{\mu\nu\rho} (A_0)_\mu
\partial_\nu (A_0)_\rho \\ &\;& +\sum_{j=\frac{1}{2}|n_1-n_2|}^{\frac{1}{2}(n_1+n_2)-1}\;
\sum_{m=-j}^{j}  \left\{\kappa \epsilon^{\mu\nu\rho} B^*_\mu(j,m)
\partial_\nu B_\rho(j,m) - \{j(j+1)-m^2\} \kappa M |B_\mu(j,m)|^2
\right\},\nonumber
\end{eqnarray}
There is no mass term for $(A_0)_\mu$ since $\hat{\alpha}_0$ commutes with both $\vp$ and
$\vp^\dagger$, and the corresponding gauge symmetry is unbroken.  For $n_1 \neq n_2$,
this is the only unbroken mode and the remaining gauge symmetry is $U(1)$. The spectrum
is similar to the maximal embedding case, except for $j$ now ranging from
$\frac{1}{2}|n_1-n_2|$ to $\frac{1}{2}(n_1+n_2)-1$. We will call it the $SU(n_1)\times
SU(n_2)$ spectrum in short. On the other hand, if $n_2 = n_1$ we have in the
"off-diagonal" sector the state $\hat{\beta}(0,0)$, which also commutes with both $\vp$
and $\vp^\dagger$. As a result, the remaining gauge symmetry becomes $SU(2)$ since the
gauge field corresponding to $\hat{\beta}(0,0)$ is complex. Except for the state
$\hat{\beta}(0,0)$, the spectrum of the "off-diagonal" sector is completely identical to
that of the "diagonal" sector, and we have now four copies of $SU(n_1)$ spectrum.

We can expand $\phi$ in a similar way:
\begin{eqnarray}
\phi &\;& = \phi_1(j_1,m_1) \hat{\alpha}_1(j_1,m_1) + \phi_2(j_2,m_2)
\hat{\alpha}_2(j_2,m_2) +\phi_0 \hat{\alpha}_0 \\ \nonumber &\;& \quad +
\sum_{j=\frac{1}{2}|n_1-n_2|}^{\frac{1}{2}(n_1+n_2)-1}\; \sum_{m=-j}^{j}\eta_1(j,m)
\hat{\beta}(j,m) + \eta_2(j,m) \hat{\beta}^\dagger(j,m) . \nonumber
\end{eqnarray}
Again, the mass spectra for $\phi_1$ and $\phi_2$ are known, and we will only give the
part for $\phi_0$ and $\eta_1, \eta_2$.
 \begin{eqnarray}
({\cal L}'_0)_{Higgs} &\;& = |\partial_\mu \phi_0|^2 - M^2 |\phi_0|^2\nonumber\\ &\;&
\quad + \sum_{j=\frac{1}{2}|n_1-n_2|}^{\frac{1}{2}(n_1+n_2)-1} \left\{|\partial_\mu
\eta_1(j,-j)|^2 - (j+1)^2 M^2|\eta_1(j,-j)|^2 \right\} \nonumber\\ &\;&   \quad +
\sum_{j=\frac{1}{2}|n_1-n_2|}^{\frac{1}{2}(n_1+n_2)-1} \left\{|\partial_\mu
\eta_2(j,j)|^2 - (j+1)^2 M^2|\eta_2(j,j)|^2 \right\} \nonumber\\ &\;& \quad +
\sum_{j=\frac{1}{2}|n_1-n_2|}^{\frac{1}{2}(n_1+n_2)-1} \sum_{m=-j+1}^{j-1}
\left\{|\partial_\mu \tilde{\eta}(j,m)|^2 - \{j(j+1)-m^2\}^2 M^2
|\tilde{\eta}(j,m)|^2\right\}.
\end{eqnarray}
Here,
\begin{eqnarray}
\tilde{\eta}(j,m) = \sqrt{\frac{\lambda(j,m-1)}{j(j+1)-m^2}}\eta_1(j,m+1) -
\sqrt{\frac{\lambda(j,m)}{j(j+1)-m^2}} \eta_2^*(j,m-1). \nonumber
\end{eqnarray}
When $n_1 \neq n_2$ the $\eta$ fields have $2n_1 n_2$ degrees of freedom. On the other
hand, if $n_2=n_1$, there are two (instead of one) more $j=0$ states, $\eta_1(0,0)$ and $
\eta_2(0,0)$, in addition to the complex neutral mode $\phi_0$.  This reflects the
remaining $SU(2)$ gauge symmetry.  Therefore, the total degrees of freedom of the $\eta$
fields become $2n_1 n_2 +2$, so that the total bosonic degrees of freedom remains to be
$4 n_1 n_2$.

Using similar expansion in $\psi$
\begin{eqnarray}
\psi &\;& = \psi_1(j_1,m_1) \hat{\alpha}_1(j_1,m_1) + \psi_2(j_2,m_2)
\hat{\alpha}_2(j_2,m_2) +\psi_0 \hat{\alpha}_0 \\ \nonumber &\;& \quad +
\sum_{j=\frac{1}{2}|n_1-n_2|}^{\frac{1}{2}(n_1+n_2)-1}\; \sum_{m=-j}^{j}\chi_1(j,m)
\hat{\beta}(j,m) + \chi_2(j,m) \hat{\beta}^\dagger(j,m), \nonumber
\end{eqnarray}
we can find the spectrum: \vfill\eject

\begin{eqnarray}
&\;& \hskip -1.5cm ({\cal L}'_0)_{Fermion} = \quad i \bar{\psi}_0\partial \hskip -0.60em
\slash\, \psi_0 - M \bar{\psi}_0 \psi_0  \nonumber \\ &\;& \qquad +
\sum_{j=\frac{1}{2}|n_1-n_2|}^{\frac{1}{2}(n_1+n_2)-1} \biggl[i
\bar{\chi}_1(j,-j)\partial \hskip -0.60em \slash\, \chi_1(j,-j) - \{ j+1\} M
\bar{\chi}_1(j,-j) \chi_1(j,-j)   \nonumber \\ &\;& \qquad\qquad\qquad\quad +
i\bar{\chi}_2(j,j)\partial \hskip -0.60em \slash\, \chi_2(j,j) - \{ j+1\} M
\bar{\chi}_2(j,j) \chi_2(j,j)  \nonumber \\ &\;& \qquad\qquad\qquad\quad +\,
i\bar{\chi}_1(j,-j+1)\partial \hskip -0.60em \slash\, \chi_1(j,-j+1) + \{ j \} M
\bar{\chi}_1(j,-j+1) \chi_1(j,-j+1) \nonumber \\ &\;& \qquad\qquad\qquad\quad +
i\bar{\chi}_2(j,j-1)\partial \hskip -0.60em \slash\, \chi_2(j,j-1) + \{ j \} M
\bar{\chi}_2(j,j-1) \chi_2(j,j-1) \biggr] \\ &\;& \qquad +
\sum_{j=\frac{1}{2}|n_1-n_2|}^{\frac{1}{2}(n_1+n_2)-1}\sum_{m=-j+ 1}^{j-1} \biggl[
i\bar{\chi}_1(j,m+1)\partial \hskip -0.60em \slash\, \chi_1(j,m+1) + \, \{ j(j+1)- m^2 \}
M \bar{\chi}_1(j,m+1) \chi_1(j,m+1)  \nonumber \\ &\;& \qquad\qquad\qquad\qquad\qquad +
i\bar{\chi}_2(j,m+1)\partial \hskip -0.60em \slash\, \chi_2(j,m+1) +\, \{ j(j+1)- m^2 \}
M \bar{\chi}_2(j,m+1) \chi_2(j,m+1) \biggr]. \nonumber \label{L0_Fermion}
\end{eqnarray}
Again, only the part for $\psi_0$, $\chi_1$, and $\chi_2$ are given. From the above
spectra, we see the neutral pair $\left\{\psi_0,\phi_0 \right\}$ and charged pairs
$\left\{B_\mu(j,j), \chi_1(j,j)\right\}$, $\left\{B_\mu(j,-j), \chi_2(j,j)\right\}$,
$\left\{\chi_1(j,-j),\eta_1(j,-j)\right\}$, $\left\{\chi_2(j,j),\eta_2(j,j)\right\}$ also
saturate the Bogomol'nyi bound and form reduced representations of the $N=2$
supersymmetry.

Now, it is quite straightforward to extend the above results to a general vacuum state
where $n = \sum_{i=1}^{k} m_i n_i$. In the i-th "diagonal" sector, we have $m_i^2$ copies
of $SU(n_i)$ spectrum.  In the "off-diagonal" sector $(i,j)$, we have $m_i m_j$ copies of
$SU(n_i)\times SU(n_j)$ spectrum. The symmetry breaking pattern in this case is $SU(n)
\to U(1)^{k-1}\times SU(m_1)\times \ldots \times SU(m_k)$. Correspondingly, we will also
have $\sum_{i=1}^{k} (m_i^2 -1) + (k-1)$ pairs of scalars and spinors with mass $M$.
Obviously, the topology in such a vacuum is $(Z_{n_1})^{m_1}
\oplus\ldots\oplus(Z_{n_k})^{m_k}$.  By comparing our results of the various phases in
$SU(3)$, $SU(4)$, and $SU(5)$ with those listed in Tables 1, 2, and 3 in Ref.
\cite{Spectra}, we check the validity of the above predictions.

In deriving our results, we rely only on the commutation relation. Thus, we should be
able to carry out similar analysis to other Lie groups.  In particular, the spectra in
Table 2 for the "maximal embedding" case of $SO(2r)$ can be obtained by the direct sum of
$SO(2r-1)$ and $SO(1)$, so that they are given by $J(J+1)-M^2$ with $J=j_1, j_2$, or $j$,
with $M = -J, -J+1,\ldots, J$ \cite{Degeneracy}. Here, $j_1 =1,3,\ldots, 2r-3$, $j_2 =0$
(which does not correspond to any physical degree of freedom), and $j = r-1$. It is can
be checked that the total number of basis vectors obtained this way does match with the
number of generators of $SO(2r)$.  We will discuss the details in an upcoming paper.

\vskip1cm

\noindent{\bf Acknowledgments} The author is indebted to G. Dunne for helpful
communication. This work is supported in part by the National Science Council of R.O.C.
under grant No. NSC89-2112-M-032-002. \vfill\eject

\vbox{ \tabskip=0pt \offinterlineskip
\hrule
\halign{\vrule#\quad&\strut\hfil#\hfil&\vrule#&\quad#\hfil&\vrule#&\quad#\hfil&\quad#\hfil&\quad#\hfil&\quad#\hfil&\quad#\hfil&\quad#\hfil&\quad\vrule#\hfil&\quad#\hfil&\quad\vrule#\tabskip=0pt\cr
&&\multispan{11}\hfil gauge masses \hfil&\cr \noalign{\hrule} &&&\hfil real
fields\quad\hfil&&\multispan{8}\hfil complex fields \hfil&\cr \noalign{\hrule} &\quad
m$\;$&&\quad 0&&1&2&3&4&\ldots &n-2&&j&\cr \noalign{\vskip -10 pt}
&j\hfill&&\omit&&\omit&\omit&\omit&\omit&\omit&\omit&&\omit& \cr
height2pt&\omit&&\omit&&\omit&\omit&\omit&\omit&\omit&\omit&&\omit& \cr \noalign{\hrule}
\noalign{\vskip -5 pt} &1\hfill&&\quad 2 && &&&&&&&1&\cr \noalign{\vskip -10 pt}
&2\hfill&&\quad 6 &&5 &&&&&&&2&\cr \noalign{\vskip -10 pt} &3\hfill&&\quad 12 &&11
&8&&&&&&3&\cr \noalign{\vskip -10 pt} &4\hfill&&\quad 20 &&19 &16&11&&&&&4&\cr
\noalign{\vskip -10 pt} &5\hfill&&\quad 30 &&29 &26&21&14&&&&5&\cr \noalign{\vskip -10
pt} &\vdots\hfill&&\quad \vdots &&\vdots &\vdots&\vdots &\vdots&$\ddots$&&&\vdots&\cr
\noalign{\vskip -10 pt} &n-1\quad\hfill&&\quad (n-1)n &&(n-1)n-1 &(n-1)n-4&(n-1)n-9
&(n-1)n-16&\ldots&3n-4&&n-1&\cr \noalign{\hrule} }} \vskip 10pt

\vbox{ \offinterlineskip
\hrule
\halign{\vrule#\quad&\strut\hfil#\hfil&\vrule#&\quad#\hfil&\vrule#&\quad#\hfil&\quad#\hfil&\quad#\hfil&\quad#\hfil&\quad#\hfil&\quad#\hfil&\quad\vrule#\hfil&\quad#\hfil&\quad\vrule#\cr
&&\multispan{11}\hfil scalar masses \hfil&\cr \noalign{\hrule} &&&\hfil real
fields\quad\hfil&&\multispan{8}\hfil complex fields \hfil&\cr \noalign{\hrule} &\quad
m$\;$&&\quad 0&&1&2&3&4&\ldots &n-2&&j&\cr \noalign{\vskip -10 pt}
&j\hfill&&\omit&&\omit&\omit&\omit&\omit&\omit&\omit&&\omit& \cr
height2pt&\omit&&\omit&&\omit&\omit&\omit&\omit&\omit&\omit&&\omit& \cr \noalign{\hrule}
\noalign{\vskip -5 pt} &1\hfill&&\quad 2 && &&&&&&&2&\cr \noalign{\vskip -10 pt}
&2\hfill&&\quad 6 &&5 &&&&&&&3&\cr \noalign{\vskip -10 pt} &3\hfill&&\quad 12 &&11
&8&&&&&&4&\cr \noalign{\vskip -10 pt} &4\hfill&&\quad 20 &&19 &16&11&&&&&5&\cr
\noalign{\vskip -10 pt} &5\hfill&&\quad 30 &&29 &26&21&14&&&&6&\cr \noalign{\vskip -10
pt} &\vdots\hfill&&\quad \vdots &&\vdots &\vdots&\vdots &\vdots&$\ddots$&&&\vdots&\cr
\noalign{\vskip -10 pt} &n-1\quad\hfill&&\quad (n-1)n &&(n-1)n-1 &(n-1)n-4&(n-1)n-9
&(n-1)n-16&\ldots&3n-4&&n&\cr \noalign{\hrule} }} \vskip 10pt

\vbox{ \offinterlineskip
\hrule
\halign{\vrule#\quad&\strut\hfil#\hfil&\vrule#&\quad#\hfil&\quad#\hfil&\quad#\hfil&\quad#\hfil&\quad#\hfil&\quad#\hfil&\quad#\hfil&\quad\vrule#\hfil&\quad#\hfil&\quad\vrule#\hfil&\quad#\hfil&\quad\vrule#\cr
&&\multispan{12}\hfil spinor masses \hfil&\cr \noalign{\hrule} &\quad m$\;$&&\quad
0&$\pm$1&$\pm$2&$\pm$3&$\pm$4&\ldots &$\pm$(n-2)&&j&&-j&\cr \noalign{\vskip -10 pt}
&j\hfill&&\omit&\omit&\omit&\omit&\omit&\omit&\omit&&\omit&&& \cr
height2pt&\omit&&\omit&\omit&\omit&\omit&\omit&\omit&\omit&&\omit&&& \cr \noalign{\hrule}
\noalign{\vskip -5 pt} &1\hfill&&2 & &&&&&&&1&&2&\cr \noalign{\vskip -10 pt} &2\hfill&&6
&5 &&&&&&&2&&3&\cr \noalign{\vskip -10 pt} &3\hfill&&12 &11 &8&&&&&&3&&4&\cr
\noalign{\vskip -10 pt} &4\hfill&&20 &19 &16&11&&&&&4&&5&\cr \noalign{\vskip -10 pt}
&5\hfill&&30 &29 &26&21&14&&&&5&&6&\cr \noalign{\vskip -10 pt} &\vdots\hfill&&\quad
\vdots &\vdots &\vdots&\vdots &\vdots&$\ddots$&&&\vdots&&\vdots&\cr \noalign{\vskip -10
pt} &n-1\quad\hfill&&(n-1)n &(n-1)n-1 &(n-1)n-4&(n-1)n-9
&(n-1)n-16&\ldots&3n-4&&n-1&&n&\cr \noalign{\hrule} }} \vskip 5pt \vbox{\small \noindent
Table.1 $SU(N)$ mass spectrum for the maximal symmetry breaking vacuum, in units of
$\frac{v^2}{\kappa}$. \hfil\break \vskip -33pt \noindent $\qquad\quad$ For $-j+1 \leq m
\leq j-1$ the gauge, spinor, and the scalar mass spectra are degenerate \hfil\break
\vskip -33pt \noindent $\qquad\quad$ forming regular representations of the $N=2$
supersymmetry.  On the other hand, \hfil\break \vskip -33pt \noindent $\qquad\quad$
$\left\{A_\mu(j,j), \psi^*(j,-j+1)\right\}$ and $\left\{\psi(j,-j),\phi(j,-j)\right\}$
saturate the Bogomol'nyi bound and \hfil\break \vskip -33pt \noindent $\qquad\quad$ form
reduced representations of the $N=2$ supersymmetry separately.} \vfill\eject

\end{document}